# Experimental demonstration of non-magnetic metamaterial cloak at microwave frequencies

Boubacar Kanté*, Dylan Germain, and André de Lustrac
Institut d'Electronique Fondamentale, Université Paris-Sud, CNRS UMR 8622, Orsay, F-91405 France
boubacar.kante@ief.u-psud.fr, andre.delustrac@ief.u-psud.fr

Metamaterials have paved the way to unprecedented control of the electromagnetic field[1,2]. The conjunction with space coordinate transformation has led to a novel "relativity inspired" approach for the control of light propagation. "Invisibility cloak" is the most fascinating proposed devices[3,4]. However, the realized structures up to now used a graded "meta-magnetic" so as to achieve the cloaking function[11]. Artificial magnetism is however still very challenging to obtain in optics despite the currently promising building blocks[13-17], not suited for optical cloaking. We report here the first experimental demonstration of non-magnetic cloak at microwave frequencies by direct mapping of the magnetic field together with the first experimental characterization of a cloak in free space configuration. The diameter of the concealed region is as big as 4.4 in wavelength units, the biggest reported experimentally so far. The principle can be scaled down to optical domain while keeping the compatibility with current nanofabrication technologies.

The interest in the emerging field of metamaterials has considerably increased during the last few years owing to their unique ability to handle both the electric and/or magnetic component of light. Most of the natural occurring materials interact with the only electric component of electromagnetic field due to the fundamental lack of magnetism in nature at high frequencies. Artificial magnetism has recently been rendered possible by means of metamaterials after the pioneering work by Pendry et al.[1]. Artificial magnetism based on Split Ring Resonators (SRRs) has thus beaten the missing magnetism in nature and has been used as constituent of the counter intuitive negative index metamaterials[2]. Based on a "general relativity" approach of electrical engineering[3,4], transformation optics has open a new route in the design of complex optical devices including electromagnetic wormholes, beam shifters and super antennas[5-8]. It just has to be told what path light must follow, and the required metamaterials parameters can be calculated (even if the result is generally very complex and unpractical).

Among the striking proposed devices is the fascinating and ancestral humans' dream of invisibility cloak. Alternative route for cloaking including plasmonic cancellation (for sub wavelength target) have also been proposed[9]. Transformation optics assigns the space squeezing to materials following the materials interpretation instead of topological interpretation[10] and led to the first invisibility cloak at microwave frequencies[11]. The device was based on a graded meta-magnetic. There is a strong motivation to translate the first demonstration to optics regarding both fundamental aspects and potential applications. A direct scaling of this first reported cloak to optics is however not possible thanks to, not only the saturation of SRRs[12] but also the very challenging resulting structure in term of nanofabrication. Alternative building blocks for optical artificial magnetism have been proposed[13-17] but none of them is suited for a cloak design. Others cloak designs for non-magnetic cloaking have been proposed[18] but have not been verified experimentally up to now due to the challenge in designing the corresponding building blocks. Instead of the complete cloak, focus has turned to broadband "under carpet cloaks"[19], simply working in reflection and experimentally verified both in microwaves[20] and infrared[21,22]. Under carpet cloaks however considerably limit cloaking potential. The original cloak, capable of restoring wave both in transmission/reflection and rendering an entire region of space invisible has not been realized in optics so far despite recent report on approximate optical invisibility device[23]. Our design and the possibility of replacing SRRs by simple metallic cut-wires[24] make true optical cloak feasible. The previously reported cloak has been measured with a polarization parallel to the cylinder axis by mapping the electric field in a guided wave setup. We report here the first experimental non-magnetic cloak with, not only the first measurement of a cloak in free space but also the biggest concealed region reported so far. The structure presented here is based on our recent proposal of a non-magnetic cloak[24] that can be designed from microwaves to optics. It is based on the electric response of SRRs instead of their magnetic response mainly used so far and thus presents SRRs as "universal atom" in the design of metamaterials. The design starts with the squeezing of space from within a cylinder of radius $b$ to within a shell of identical external radius $b$ and internal radius $a$. From the invariance of Maxwell equations, the space compression can be assigned to material parameters i.e. dielectric permittivity ($\varepsilon$) and magnetic permeability ($\mu$) leading to a rather complicated set of anisotropic and inhomogeneous materials. The full set of parameters is given by:

$$\varepsilon_r = \mu_r = \frac{r-a}{r}, \quad \varepsilon_\theta = \mu_\theta = \frac{r}{r-a}, \quad \varepsilon_z = \mu_z = \left(\frac{b}{b-a}\right)^2 \frac{r-a}{r} \quad (1)$$

However, for a given polarization, only some tensors components are relevant. Considering a non-magnetic cloak, the set of parameters to fulfill reported in ref. 18 and 24 is the following:

$$\mu_z = 1, \quad \varepsilon_\theta = \left(\frac{b}{b-a}\right)^2, \quad \varepsilon_r = \left(\frac{b}{b-a}\right)^2 \left(\frac{r-a}{r}\right)^2 \quad (2)$$

This reduced set of parameters holds for a polarization perpendicular to the cylinder axis and satisfies the dispersion relation, but not the equality of the wave impedance between the vacuum and the cloak. A non-zero reflection is then predictable. The level of power reflection $R_p$ can be estimated by[18]:

$$R_p = |(1-Z)/(1+Z)|^2 = \left[\left(\frac{a}{b}\right) \Big/ \left(2 - \left(\frac{a}{b}\right)\right)\right]^2 \quad (3)$$

The fabricated anisotropic metamaterials is produced from copper SRRs embedded in a dielectric host, a commercial resin with a permittivity of 2.75. In our implementation, the radial permittivity profile is designed using the electric response of SRRs[24] by locally changing the dimension of resonators, actually only the SRRs gap size. The typical unit cell and the effective parameters[25] of discrete SRRs at the design frequency (11 GHz) are presented in Fig. 1.

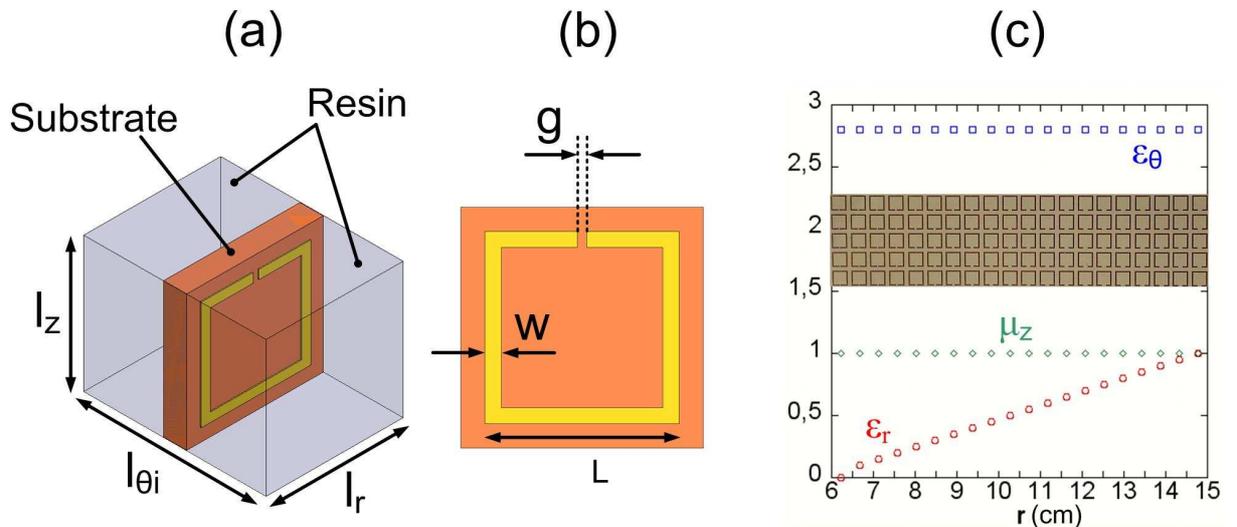

Fig.1: (a) Unit cell. (b) The dimensions of a typical square SRR are: L=3.6 mm, w=0.3 mm and copper thickness t=35 µm. The SRRs gap g and $l_{\theta i}$, are the only varying parameters. $l_{\theta i}$ linearly decreases from the outer to the inner boundary of the cloak (see supplementary information). (c) Picture of a radial stripes and corresponding relevant effective parameters.

Each resonator is assumed to have the effective parameters of an infinite array of identical elements. While SRRs (embedded in the host medium) are used to achieve the radial variation of the permittivity function, the azimutal permittivity is mainly implemented by the permittivity of the host medium itself since SRRs have no electric response in this direction.
The realized cloak is comprised of 15700 elementary SRRs. The cylindrical shell is divided in 20 annular regions of equal thickness ($l_r$=4.5 mm) with a linear radial variation of the permittivity from 0 to 1 (from the inner to the outer boundary of the cloak) and 157 stripes separated by an angle of about 2.3° (Fig.2 (a)). The inner and outer radius of the cloak are *a*=6 cm, *b*=15 cm and the cloak height is 2.25 cm corresponding to 5×$l_z$ i.e. 5 SRR layers. For this set of parameters the reflection coefficient $R_p$ is very weak, equal to 0.0625. The SRRs within a given annular region are identical and designed to have the proper local radial permittivity.
The host medium, a commercially available resin is an important design component (closely linked to $\varepsilon_\theta$). Its permittivity has been measured and found to be equal to $\varepsilon_{resin}$= 2.75. The SRRs have been printed on a dielectric substrate (as seen in the picture of Fig.1 (c)) with a permittivity close to the resin's one. We chose RO3003 with $\varepsilon_{substrate}$= 3±0.04 and a dielectric loss tangent at 11 GHz of about 0.0013. The SRRs have been chemically etched on this substrate. The design process as well as the dimensions of the cloak elements can be found in Supplementary Information. The 157 stripes were arranged in a moulded watertight polymeric matrix designed accordingly (Fig. 2 (a)). Then the liquid resin was run in the silicon mould. After solidification, the cloak was removed from the mould and can be easily manipulated (Fig. 2 (b)).

In contrast with previously reported structures, the measurements are performed in free space and not in a waveguide (Fig. 2 (c)). A loop antenna, consisting of a circular coil made of the inner conductor of a SMA cable has been designed to map the magnetic field ($H_z$). The magnetic field is output from the X-band horn antenna. Both antennas are connected to an Agilent 8722ES Vectorial Network Analyzer. The loop antenna position can be controlled via an automated Labview program over a surface of 40 cm*40 cm and getting for each spatial position of the loop antenna the complex (magnitude and phase) scattering parameters. It is important to notice that the cloak's height is only 2.25 cm. An absorbing window of comparable height has thus been placed at the horn antenna output to limit the leaking of waves above and below the cloak so that most of the energy travels into the cloak. The experimental setup can be seen on Fig. 2 (c). Since the structure is filled by the resin, it is difficult to access the internal field. Instead, the bottom surface of the cloak (see Fig. 2 (c)) has been scanned taking profit of the continuity of field at this boundary in quasi-contact mode.

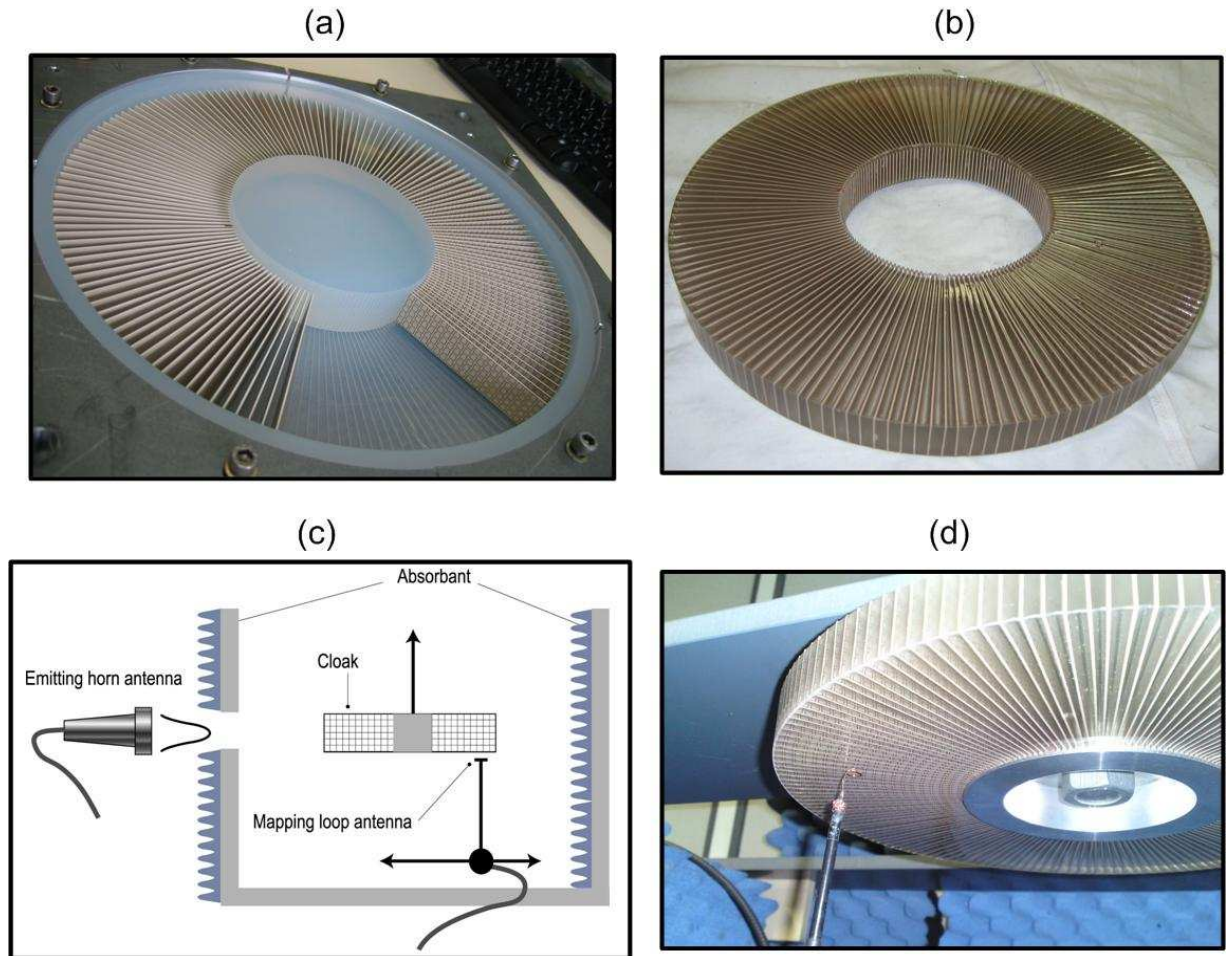

Fig. 2: (a) Assembling of the invisibility cloak from metamaterials based SRRs stripes in the silicone mould. (b) Realized invisibility cloak (c) Two dimensional view of the experimental setup. (d) Picture of a portion of the experimental setup with the loop antenna mapping the magnetic field at the bottom surface of the cloak.

To evaluate the performance of the cloak, three measurements are done with a scanning step size of 2 mm. The first measurement maps the magnetic field of the free space radiation from the horn antenna (Fig. 3 (a)). The second and third measurements use a metallic cylinder alone (diameter 12 cm) (Fig. 3 (b)) and surrounded with the cloak (outer diameter 30 cm) (Fig. 3 (d)). The results are presented in Figs. 3 (real part of the complex transmission). The quasi-cylindrical wave output from the horn antenna is nicely resolved in our measurement (Fig. 3 (a)). In presence of the metallic cylinder, the scattering and shadowing effects can be clearly observed in Fig. 3 (b) as well as interferences between the incident and reflected beams. Fig. 3 (d) shows that in the presence of the cloak, the shadowing effect of the metallic cylinder is suppressed and the wave fronts are maintained thus demonstrating the cloaking effect. For comparison, simulation result using commercial finite element code (Comsol Multiphysics) for a cloak with the reduced parameters of equations (2) is reported in Fig. 3 (c). The very good agreement with measurement can be noticed. More importantly, the bending and redirection of quasi-cylindrical wave fronts inside the cloak can nicely be observed as a change in the radius of the horn antenna waves fronts in Fig. 3 (d).

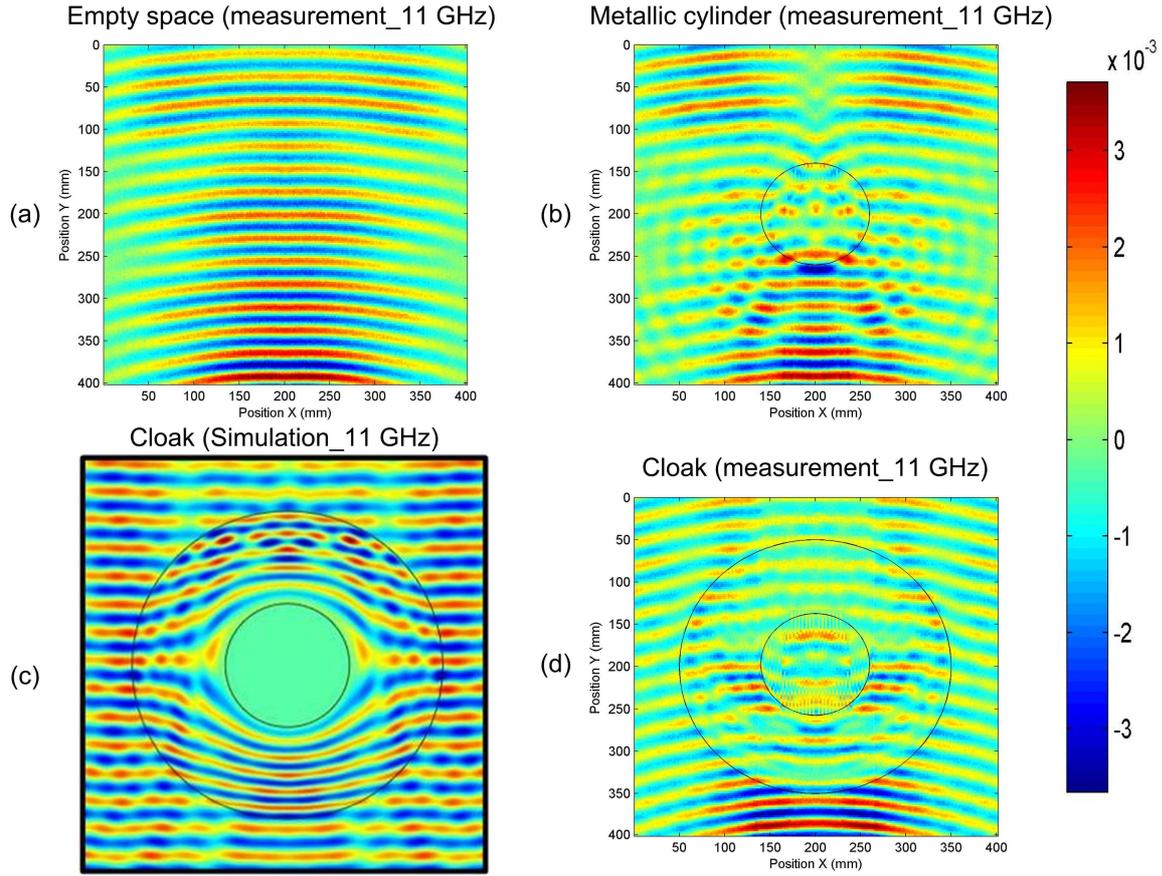

Fig. 3: Real part of the measured magnetic field output from the horn antenna in free space (a) with the metallic cylinder alone (b) and with the cloak surrounding the metallic cylinder (d). Finite element simulation exciting the cloak by plane wave (Comsol Multiphysics) with the reduced set of parameters presented in equations (1) is reported for comparison (c). In all cases, the 11 GHz wave travels from bottom to top.

The non-magnetic cloak demonstrated here works for a magnetic field parallel to the cylinder axis. Associating an electric resonator (our structure or electrical resonators in ref. 20) with the one reported by Schurig *et al.*[11] could lead to a polarization independent cloak. It should however be recognize that the corresponding structure, feasible from microwave to terahertz will be very difficult to realize in the near infrared or optical domain despite recent progress in three dimensional metamaterials fabrication[16]. However, the SRR can be adiabatically transformed into a single cut-wire[26]. Such cut-wires structure has shown to be comparable to SRRs with regard to their electrical response[27]. The non-magnetic cloak reported here can thus be scaled down to infrared and visible by simply replacing the SRRs by cut-wires[24]. The corresponding design would be compatible with current nanofabrication thin film deposition and processing.

We have reported the first experimental demonstration of non-magnetic cloak at microwave frequencies. The measurements have been performed in free space with the biggest concealed region reported so far (4.4 in wavelength units). The electric response of split ring resonators has been used to achieve the cloaking function for a polarization perpendicular to the cylinder axis. Therefore, SRRs can be presented as "universal atom" in the design of metamaterials with their magnetic and electric responses that can be used separately or in conjunction. It has also been shown that our cloak can be scaled down to optics while keeping its compatibility with nanofabrication techniques by replacing the SRRs with simple metallic cut-wires.

The authors thank J.-M. Lourtioz and P. Crozat for fruitful discussion, T. Lepetit for help in measuring the resin and technical support from W. D. de Marcillac, F. Delgehier and A. Charrier (all from IEF, Univ. Paris-Sud 11/CNRS).

References :

1. J. B. Pendry, A. J. Holden, D. J. Robbins and W. J. Stewart, IEEE Trans. Microwave Theory Tech. **47**, 2075-2084 (1999).
2. D. R. Smith, W. J. Padilla, D. C. Vier, S .C. Nemat-Nasser and S. Schultz, Phys. Rev. Lett. **84**, 4184-4187 (2000).
3. U. Leonhardt, Science **312**, 1777-1780 (2006).
4. J. B. Pendry, D. Schurig, and D. R. Smith, Science **312**, 1780-1782 (2006).
5. A. Greenleaf, Y. Kurylev, M. Lassas, G. Uhlmann, Phys. Rev. Lett. **99**, 183901 (2007).
6. M. Rahm, S. A. Cummer, D. Schurig, J. B. Pendry, D. R. Smith, Phys. Rev. Lett. **100**, 063903 (2008).
7. U. Leonhardt, and T. Tyc, New. J. Phys. **10**, 115026 (2008).
8. P. H. Tichit, S. N. Burokur, A. de Lustrac, J. Appl. Phys. **105**, 104912 (2009).
9. A. Alu, N. Engheta, Phys. Rev. E **72**, 016623 (2005).
10. D. Schurig, J. B. Pendry, and D. R. Smith, Opt. Express **14**, 9794-9803 (2006).
11. D. Schurig, J. J. Mock, B. J. Justice, S. A. Cummer, J. B. Pendry, A. F. Starr, and D. R. Smith, Science **314**, 977-980 (2006).
12. S. O'Brien, J. B. Pendry, J. Phys. Condens. Matter **14**, 6383-6394 (2002).
13. A. N. Grigorenko, A. K. Geim, H. F. Gleeson, Y. Zhang, A. A. Firsov, I. H. Khrushchev and J. Petrovic, Nature (London) **438**, 335-338 (2005).
14. V. M. Shalaev, W. Cai, U. K. Chettiar, H. K. Yuan, A. K. Sarychev, V. P. Drachev and A. V. Kildishev, Opt. Lett. **30,** 3356-3358 (2005).
15. J. Zhou, L. Zhang, G. Tuttle, T. Koschny, and C. M. Soukoulis, Phys. Rev. B **73**, 041101(R) (2006).
16. N. Liu, H. Guo, L. Fu, S. Kaiser, H. Schweizer and H. Giessen, Nature Mater. **7**, 31-37 (2008).
17. B. Kanté, S. N. Burokur, A. Sellier, A. de Lustrac and J.-M. Lourtioz, Phys. Rev. B **79**, 075121 (2009).
18. W. Cai, U. K. Chettiar, A. V. Kildishev, and V. M. Shalaev, Nature Photonics **1**, 224-227 (2007).
19. J. Li and J. B. Pendry, Phys. Rev. Lett. **101**, 203901 (2008).
20. R. Liu, C. Ji, J. J. Mock, J. Y. Chin, T. J. Cui, D. R. Smith, Science **323**, 366-369 (2009).
21. J. Valentine, J. Li, T. Zentgraf, G. Bartal, and X. Zhang, Nature Mater. **8**, 568-571 (2009).
22. L. H. Gabrielli, J. Cardenas, C. B. Poitras and M. Lipson, "Cloaking at optical frequencies". Preprint at <http://arxiv.org/abs/0904.3508> (2009).
23. I. I. Smolyaninov, Y. J. Hung, and C. C. Davis, Opt. Lett. **33**, 1342-1344 (2008).
24. B. Kanté, A. de Lustrac, J.-M. Lourtioz, S. N. Burokur, Opt. Express **16**, 9191 (2008).
25. A. M. Nicolson, and G. F. Ross, IEEE Trans. Instrum. Meas. **19**, 377-382 (1970).
26. C. Rockstuhl, F. Lederer, C. Etrich, T. Zentgraf, J. Kuhl and H. Giessen, Opt. Express **14**, 8827 (2006).
27. B. Kanté, A. de Lustrac, J.-M. Lourtioz, F. Gadot, Opt. Express **16**, 6774 (2008).